\documentclass[11pt,a4paper]{article}

\usepackage{epsfig,amsmath,amssymb,cite}

\begin{document}

\title{Gravitational lensing of transient neutrino sources by black holes}

\author{Ernesto F. Eiroa$^{1,2,}$\thanks{
e-mail: eiroa@iafe.uba.ar}, Gustavo E. Romero$^{3,4,}$\thanks{
e-mail: romero@iar-conicet.gov.ar, Member of CONICET} \\
{\small $^1$ Departamento de F\'{\i}sica, Facultad de Ciencias Exactas y Naturales, Universidad de Buenos Aires, }\\
{\small Ciudad Universitaria Pab. I, 1428, Buenos Aires, Argentina}\\
{\small $^2$ Instituto de Astronom\'{\i}a y F\'{\i}sica del Espacio, C.C.
67, Suc. 28, 1428, Buenos Aires, Argentina}\\
{\small $^3$ Instituto Argentino de Radioastronom\'{\i}a (IAR-CONICET), C.C.5, 1894 Villa Elisa, Buenos Aires, Argentina}\\
{\small $^4$ Facultad de Ciencias Astron\'omicas y Geof\'{\i}sicas, Universidad Nacional de La Plata, }\\
{\small Paseo del Bosque, 1900 La Plata, Argentina}}

\maketitle

\begin{abstract}
In this work we study gravitational lensing of neutrinos by Schwarzschild black holes. In particular, we analyze the case of a neutrino transient source associated with a gamma-ray burst lensed by a supermassive black hole located at the center of an interposed galaxy. We show that the primary and secondary images have an angular separation beyond the resolution of forthcoming km-scale detectors, but the signals from each image have time delays between them that in most cases are longer than the typical duration of the intrinsic events. In this way, the signal from different images can be detected as separate events coming from the very same location in the sky. This would render an event that otherwise might have had a low signal-to-noise ratio a clear detection, since the probability of a repetition of a signal from the same direction is negligible. The relativistic images are so faint and proximate that are beyond the sensitivity and resolution of the next-generation instruments.
\end{abstract}



\section{Introduction}

Gravitational lensing of photons by black holes has received great attention in the last few years, mainly due to the increasing evidence of the presence of supermassive black holes at the center of galaxies. Theoretical studies of black hole lenses, both numerically and analytically, were made with Schwarzschild \cite{virbha1, bozza1, schw}, Reissner--Nordstr\"{o}m \cite{eiroto}, general spherically symmetric \cite{bozza2, bozman1} and rotating \cite{bozza3, rotating} geometries, and also for black holes coming from alternative theories \cite{alternative} or braneworld cosmologies \cite{braneworld}. Even naked singularities were considered as lenses \cite{nakedsing}. Photons (or null mass particles) passing close enough to the photon sphere of the lens will have large deflection angles, and they can even make one or more turns around the deflector before reaching the observer. By this mechanism, two infinite sets of strong deflection images, one at each side of the lens, are produced. The presence of images with large deflection angles is not a new fact, since they were obtained already in 1959 for the Scharzschild spacetime \cite{darwin}. The analytical study of these images is more simple if one adopts the strong deflection limit, which consists in a logarithmic approximation of the deflection angle, first obtained for the Schwarzschild metric \cite{darwin}, revisited by several authors \cite{old, bozza1},  extended to Reissner--Nordstr\"{o}m geometry \cite{eiroto}, to general spherically symmetric spacetimes \cite{bozza2} and to Kerr metric \cite{bozza3}.  For some lensing configurations two weak deflection images are also obtained, which are analyzed by making a first order Taylor expansion of the deflection angle (weak deflection limit), as it is usually done for more standard astrophysical objects, such as stars and galaxies (see, e.g., \cite{schneider}). Intermediate cases can be treated analytically by perturbative \cite{perturba} or variational methods \cite{amore}. A special configuration, where no weak deflection images are present, is when the source is in front of the lens instead of behind it, which is called retrolensing \cite{retrolens}. Recently, the strong deflection limit was extended to include sources very close to the black hole \cite{bozza4}.\\

Lensing of neutrinos have been previously studied by other authors. In Ref. \cite {escribano}, gravitational lensing of neutrinos by stars and galaxies was analyzed, and in Ref. \cite{mena}, the lensing effects of supernova neutrinos by the Galactic center black hole was considered, in the weak deflection limit. However, perhaps the most interesting cosmological sources of neutrinos from the point of view of lensing are transients associated with gamma-ray bursts (GRBs). It is expected that proton-photon interactions during the GRB will result into copious photopion production and hence neutrinos would be generated from the decay of charged pions and muons (e.g. \cite{neutrinos, meszaros, dermer, murase}). Since GRBs occur frequently, say once per day, and can be detected at gamma-rays by SWIFT satellite and then the follow up of the afterglows usually allows the identification of the host galaxy and the corresponding redshift (see, e.g., \cite{meszaros2}), they are outstanding candidates for lensing produced by massive black holes in the center of interposed galaxies. \\

We notice, however, that in the collapsar scenario for long GRBs \cite{woosley1, woosley2} the jet not always is expected to be able to make its way through the star, so no observable gamma-ray emission would result in such cases \cite{meszaros}. Nonetheless, the neutrino emission might be important. If the event is lensed, the neutrino signal should repeat and hence be identified, despite the absence of electromagnetic counterparts. \\

In this letter we investigate gravitational lensing of neutrinos by Schwarzschild black holes. We pay special attention to neutrino transients lensed by supermassive black holes located at the center of galaxies. In Section \ref{pm} we present the expressions that give the positions and magnifications of the weak and strong deflection images, and in Section \ref{td} we calculate the time delays between the arrival signals. Then, in Section \ref{lnb}, we calculate the specific time delays produced by some interposed supermassive black holes for neutrino transient at a distance of $\sim 10^{28}$ cm. Finally, in Section \ref{conc}, a brief summary and the conclusions are presented.

\section{Positions and magnifications of the images}\label{pm} 

Neutrinos have zero or negligible mass, so we assume that they follow null geodesics as photons do. We consider a point source of neutrinos, with angular diameter distance $D_{\rm{os}}$ to the observer, behind a Schwarzschild black hole lens, placed at an angular diameter distance $D_{\rm{ol}}$. The angular diameter distance between the lens and the source is dubbed $D_{\rm{ls}}$. The optical axis is defined by the line that joins the observer with the deflector. The distances are very large compared to the Schwarzschild radius of the black hole and the angles are measured from the observer. We restrict our analysis to high alignment, which is more interesting from an astrophysical point of view, since the images are more prominent. Then the angular position of the source $\beta $, taken positive here, is small. For this configuration, we have two weak deflection images and two infinite sets of strong deflection (also called relativistic \cite{virbha1}) images. Neutrinos with closest approach distance $r_{0}$ much larger than the photon sphere radius $r_{\rm{ps}}=3MG/c^2$, which corresponds to the unstable circular orbit around the black hole\footnote{For a complete study of the photon sphere in a spherically symmetric geometry see \cite{atkinson}; and for a general definition of the photon surface in an arbitrary spacetime see \cite{claudel}.}, will have a small deflection angle $\alpha$, which can be approximated to first order in $1/r_{0}$ by $\alpha= 4GM/(c^{2}r_{0})$ (weak deflection limit). Within this approximation, the lens equation has the form \cite{schneider}
\begin{equation}
\beta =\theta-\frac{\theta_{\rm{E}}^{2}}{\theta},
\label{pm1} 
\end{equation}
where $\theta $ is the angular position of the image and $\theta_{\rm{E}}$ is the angular Einstein radius, given by
\begin{equation}
\theta_{\rm{E}}=\sqrt{\frac{2R_{\rm{s}}D_{\rm{ls}}}{D_{\rm{ol}}D_{\rm{os}}}},
\label{pm2} 
\end{equation} 
with $R_{\rm{s}}=2MG/c^{2}$ the Schwarzschild radius of the lens. The lens equation has two solutions:
\begin{equation}
\theta_{\rm{p},s}=\frac{1}{2}\left( \beta\pm\sqrt{\beta^{2}+4\theta_{\rm{E}}^{2}}\right),
\label{pm3} 
\end{equation} 
that give the positions of the primary (upper sign) and the secondary (lower sign) images. The primary image lies inside the Einstein radius and the secondary image outside. When $\beta =0$, instead of two images, an Einstein ring with radius $\theta_{\rm{E}}$ is obtained. Another important aspect is the magnification of the images, defined as the ratio between the observed and intrinsic fluxes of the source. As a consequence of the Liouville theorem in curved spacetimes \cite{misner}, gravitational lensing preserves surface brightness for neutrinos and photons, so the magnifications of the images are given by the ratio of the solid angles subtended by the images and the source, which result in \cite{schneider}:
\begin{equation}
\mu _{\rm{p},s}=\frac{1}{4}\left( \frac{\beta}{\sqrt{\beta^{2}+4\theta_{\rm{E}}^{2}}}+
\frac{\sqrt{\beta^{2}+4\theta_{\rm{E}}^{2}}}{\beta}\pm 2 \right),
\label{pm4} 
\end{equation}
where the plus sign corresponds to the primary image and the minus sign to the secondary one. If the position of the source $\beta$ is close to zero, the magnifications of both images are large. If $\beta =0$ the approximation of point source breaks down and the magnifications become infinite. It is not difficult to see that $\mu _{\rm{p}}>1$ for all $\beta $, and $\mu _{\rm{s}}>1$ only if $\beta /\theta_{\rm{E}}< \sqrt{(3\sqrt{2}-4)/2}\approx 0.35$. When $\beta /\theta_{\rm{E}}$ is large we have that $\mu _{\rm{p}}\approx 1$ and $\mu _{\rm{s}}\approx 0$.\\

Besides the weak deflection images, two infinite sets of relativistic images are formed by neutrinos that make one or more loops, in both directions of winding, around the black hole lens. For high alignment, the deflection angle corresponding to the relativistic images is close to an even number of $\pi$, $\alpha=\pm (2n\pi + \Delta\alpha_{n})$ with $0<\Delta\alpha_{n}\ll 1$, the upper sign corresponding to one set of images and the lower one to the other set. The other angles involved are small, then the lens equation \cite{virbha1}\footnote{Eq. (\ref{pm5a}) is valid for asymptotically flat spacetimes, with the source and the observer in the flat region; for more general lens equations see \cite{lensequa}.} 
\begin{equation}
\tan \beta =\tan \theta -\frac{D_{\mathrm{ls}}}{D_{\mathrm{os}}}\left( \tan
\theta + \tan(\alpha -\theta )\right),
\label{pm5a}
\end{equation}
takes the form \cite{bozza1, bozza2}
\begin{equation}
\beta=\theta \mp \frac{D_{\rm{ls}}}{D_{\rm{os}}}\Delta\alpha_{n}.
\label{pm5b}
\end{equation}
In the strong deflection limit, i.e. for trajectories passing close to the photon sphere of the black hole, the deflection angle can be approximated by a logarithmic function of the impact parameter $b$, defined as the perpendicular distance from the deflector to the asymptotic path at infinite. For the Schwarzschild geometry, it can be shown that \cite{darwin, bozza1, bozza2}
\begin{equation}
\alpha =\pm \left[ -c_{1}\ln \left( \frac{b}{b_{\rm{ps}}}-1\right)+ c_{2}\right] +O(b-b_{\rm{ps}}),
\label{pm6} 
\end{equation} 
with $c_{1}=1$, $c_{2}=\ln [216(7-4\sqrt{3})]-\pi$ and $b_{\rm{ps}}=\sqrt{3}r_{\rm{ps}} =3\sqrt{3}R_{\rm{s}}/2$ the critical impact parameter. Neutrinos with impact parameter smaller than the critical value will spiral inside the photon sphere into the black hole, not reaching the observer, and those with $b$ larger than $b_{\rm{ps}}$ will make one or more outward turns outside the photon sphere, finally getting to the observer. As in the case of photons, using that $b=\sin D_{\rm{ol}} \theta \approx \theta D_{\rm{ol}}$, inverting Eq. (\ref{pm6}) and Taylor expanding it around $\alpha =2n\pi$ to obtain $\Delta\alpha_{n}$, then replacing the result in the lens equation (\ref{pm5b}) and finally inverting it, the positions of the relativistic images can be approximated (keeping only the lower order terms) by \cite{bozza1, bozza2}:
\begin{equation}
\theta_{n}=\pm \theta_{n}^{\rm{E}}+\frac{D_{\rm{os}}b_{\rm{ps}}}{D_{\rm{ls}}D_{\rm{ol}}c_{1}}e_{n}\beta,
\label{pm7} 
\end{equation} 
where
\begin{equation}
 e_{n}=e^{(c_{2}-2n\pi)/c_{1}},
\nonumber
\end{equation} 
and
\begin{equation}
\theta_{n}^{\rm{E}}=\frac{b_{\rm{ps}}}{D_{\rm{ol}}}\left( 1- \frac{D_{\rm{os}}b_{\rm{ps}}}{D_{\rm{ls}}D_{\rm{ol}}c_{1}}e_{n}\right)(1+e_{n}),
\label{pm8} 
\end{equation} 
is the $n$-th relativistic Einstein ring radius. For perfect alignment an infinite sequence of Einstein rings is obtained instead of point images. With the same considerations given above for the weak deflection images, the magnification of the $n$-th image has the same expression that was found previously for photons \cite{bozza1, bozza2}:
\begin{equation}
\mu_{n}=\frac{1}{\beta}\frac{b_{\rm{ps}}^2D_{\rm{os}}}{D_{\rm{ol}}^2 D_{\rm{ls}} c_{1}}(1+e_{n})e_{n},
\label{pm9} 
\end{equation} 
for both sets of relativistic images. The first image ($n=1$) is the strongest one and the others have magnifications that decrease exponentially with $n$. For a given source angle $\beta$, the relativistic images are very faint compared with the weak deflection ones\footnote{For example, if $\beta /\theta_{\rm{E}}\ll 1$  we have that $\mu_{1}/\mu_{\rm{p}}\propto (R_{\rm{S}}/D_{\rm{ol}})^{3/2}$, which is usually a very small number.}.

\section{Time delays}\label{td} 

Neutrinos that form distinct images take different paths, resulting in time delays between the images. Considering again that neutrinos follow null geodesics as photons do, the time delay between the primary and the secondary images is given by \cite{schneider}:
\begin{equation}
\Delta t_{\rm{p,s}}=\frac{2R_{\rm{s}}}{c}(1+z_{\rm{d}})\left( \frac{\theta_{\rm{s}}^{2}-\theta_{\rm{p}}^{2}}{2|\theta_{\rm{p}}\theta_{\rm{s}}|} +\ln \left|\frac{\theta_{\rm{s}}}{\theta_{\rm{p}}}\right|\right),
\label{td1}
\end{equation}
where $z_{\rm{d}}$ is the redshift of the deflector. The last equation can be written in the form
\begin{equation}
\Delta t_{\rm{p,s}}=\frac{2R_{\rm{s}}}{c}(1+z_{\rm{d}})\left( \frac{-\beta \sqrt{\beta^{2}+4\theta_{\rm{E}}^{2}}}{2\theta_{\rm{E}}^{2}}+\ln \left| \frac{\beta -\sqrt{\beta^{2}+4\theta_{\rm{E}}^{2}}}{\beta +\sqrt{\beta^{2}+4\theta_{\rm{E}}^{2}}}\right|\right).
\label{td2}
\end{equation}
When $\beta =0$ there is no time delay. Large time delays can be obtained if $\beta /\theta_{\rm{E}}\gg 1$, but in this case the magnification of the primary image is close to one and the secondary image is very faint. The optimal situation for a variable source is when $\beta /\theta_{\rm{E}}$ is small enough to have large magnifications of both images, but not too close to zero, so the time delay can be longer than the typical time scale of the transient source.\\

In the case of relativistic images, the time delay between the images formed at the same side of the lens is given by \cite{bozman1}\footnote{The expressions from Ref. \cite{bozman1} have been rewritten here using physical units, adding the cosmological factor $1+z_{d}$ and expanding them to first order in the source position angle (measured from the observer instead of from the source).}:
\begin{equation}
\Delta t^{\rm{s}}_{n,m}=\frac{b_{\rm{ps}}}{c}(1+z_{\rm{d}})\left[ 2\pi (n-m)+2\sqrt{2}(w_{m}-w_{n})\pm \dfrac{\sqrt{2}D_{\rm{os}}(w_{m}-w_{n})}{c_{1}D_{\rm{ls}}}\beta \right] ,
\label{td3}
\end{equation}
where
\begin{equation}
w_{k}=e^{(c_{2}-2k\pi)/(2c_{1})},
\nonumber
\end{equation}
and the upper/lower sign corresponds if both images are on the same/opposite side of the source. For the images at the opposite side of the lens we have \cite{bozman1}:
\begin{equation}
\Delta t^{\rm{o}}_{n,m}=\frac{b_{\rm{ps}}}{c}(1+z_{\rm{d}})\left[ 2\pi (n-m)+2\sqrt{2}(w_{m}-w_{n})+\left(  \dfrac{\sqrt{2}D_{\rm{os}}(w_{m}+w_{n})}{c_{1}D_{\rm{ls}}}-\frac{2D_{\rm{os}}}{D_{\rm{ls}}}\right) \beta \right],
\label{td4}
\end{equation}
where the image with winding number $n$ is on the same side of the source and the other one on the opposite side. The first term in Eqs. (\ref{td3}) and (\ref{td4}) is by large the most important one \cite{bozman1}. The time delays between the relativistic images are longer than the time delay between the primary and the secondary images.

\section{Lensing of neutrino transients}\label{lnb} 

\begin{table}[t]
\begin{center}
\begin{tabular}{|c|c|c|c|c|}
\hline
Galaxy & Black hole mass ($M_{\odot}$) & Distance (Mpc) & $\theta_{\rm{E}}$ (arcsec) & $\Delta t_{\rm{s,p}}$ (s)\\
\hline
Milky Way & $2.8 \times 10^{6}$ & $0.0085$ & $1.6$  & $11$ \\
\hline
NGC0224   & $ 3.0\times 10^{7}$ & $0.7$ & $0.6$  & $1.2\times 10^{2}$\\
\hline
NGC3115   & $ 2.0\times 10^{9}$ & $8.4$ & $1.4$  & $7.9\times 10^{3}$\\
\hline
NGC3377   & $ 1.8\times 10^{8}$ & $9.9$ & $0.4$  & $7.1\times 10^{2}$\\
\hline
NGC4486B   & $ 5.7\times 10^{8}$ & $15.3$ & $0.5$  & $2.2\times 10^{3}$\\
\hline
NGC4486   & $ 3.3\times 10^{9}$ & $15.3$ & $1.3$ & $1.3\times 10^{4}$\\
\hline
NGC4261   & $ 4.5\times 10^{8}$ & $27.4$ & $0.4$ & $1.8\times 10^{3}$\\
\hline
NGC7052   & $ 3.3\times 10^{8}$ & $58.7$ & $0.2$ & $1.3\times 10^{3}$\\
\hline
\end{tabular}
\end{center}
\caption{Time delays between the weak deflection images of neutrino burst sources at a distance of $10^{28}$ cm. The lenses are supermassive black holes at the center of the galaxies indicated. The Schwarzschild geometry was adopted to model the black holes. The source angular position is $\beta =0.1 \, \theta_{\rm{E}}$, with $\theta_{\rm{E}}$ the angular Einstein radius. In this case, the angular positions of the primary and the secondary images are $\theta_{\rm{p}}=1.05 \, \theta_{\rm{E}}$ and $\theta_{\rm{s}}=-0.95 \, \theta_{\rm{E}}$, while their respective magnifications are $\mu_{\rm{p}}=5.5$ and $\mu_{\rm{s}}=4.5$.}
\label{table1}
\end{table}

\begin{table}[t]
\begin{center}
\begin{tabular}{|c|c|c|c|c|}
\hline
Galaxy & Black hole mass ($M_{\odot}$) & Distance (Mpc) & $\theta_{\rm{E}}$ (arcsec) & $\Delta t_{\rm{s,p}}$ (s)\\
\hline
Milky Way & $2.8 \times 10^{6}$ & $0.0085$ & 1.6 & $55$\\
\hline
NGC0224   & $ 3.0\times 10^{7}$ & $0.7$ &  0.6 & $6.0\times 10^{2}$\\
\hline
NGC3115   & $ 2.0\times 10^{9}$ & $8.4$ & 1.4  & $4.0\times 10^{4}$\\
\hline
NGC3377   & $ 1.8\times 10^{8}$ & $9.9$ & 0.4 & $3.6\times 10^{3}$\\
\hline
NGC4486B  & $ 5.7\times 10^{8}$ & $15.3$ & 0.5  & $1.1\times 10^{4}$\\
\hline
NGC4486   & $ 3.3\times 10^{9}$ & $15.3$ & 1.3  & $6.6\times 10^{4}$\\
\hline
NGC4261   & $ 4.5\times 10^{8}$ & $27.4$ & 0.4  & $8.9\times 10^{3}$\\
\hline
NGC7052   & $ 3.3\times 10^{8}$ & $58.7$ & $0.2$ & $6.6\times 10^{3}$\\
\hline
\end{tabular}
\end{center}
\caption{Time delays between the weak deflection images of neutrino burst sources at a distance of $10^{28}$ cm. The lenses are supermassive black holes at the center of the galaxies indicated. The Schwarzschild geometry was adopted to model the black holes. The source angular position is $\beta =0.5 \, \theta_{\rm{E}}$, with $\theta_{\rm{E}}$ the angular Einstein radius. In this case, the angular positions of the primary and the secondary images are $\theta_{\rm{p}}=1.28 \, \theta_{\rm{E}}$ and $\theta_{\rm{s}}=-0.78 \, \theta_{\rm{E}}$, while their respective magnifications are $\mu_{\rm{p}}=1.6$ and $\mu_{\rm{s}}=0.6$.}
\label{table2}
\end{table}

The angular resolution of the primary and secondary images is beyond the capability of current and near future neutrino detectors, which is of the order of one tenth of a degree, but the temporal resolution of the images of individual transient events, which have typical durations in the range of $\sim$10 s to 100 s for long GRBs \cite{meszaros2}, is possible. As an example of this, we consider neutrino transients acting as possible sources situated at distances of the order of $10^{28}$ cm, with supermassive black holes at the center of interposed galaxies as lenses. Some results of our calculations for specific cases of lenses in the local universe ($z_{\rm{d}}\sim 0$) are shown in Tables \ref{table1} and \ref{table2}, with the masses and distances taken from Ref. \cite{richstone}. We see that the separation between the primary and secondary images of neutrino transients is of the order of a second of arc, so they cannot be resolved. For suitable values of the parameters involved, the weak deflection images can be both magnified several times, with time delays of $10^2 - 10^4$ s, larger than the intrinsic time of variation of the sources. \\

If one fixes $\beta /\theta_{\rm{E}}$ to obtain from Eq. (\ref{pm4}) the desired values of magnification of the images, it is clear from Eq. (\ref{td2}) that the time delay increases linearly with the redshift of the lens. Then, with a typical source with redshift $z_{\rm{s}}\sim 1$, the values of time delays between the primary and the secondary images for lenses closer to the neutrino sources can be up to twice of those obtained in Tables \ref{table1} and \ref{table2}. But far lenses require better alignment to have large magnifications because $\theta _{\rm{E}}$ decreases with the distance to the lens.\\

Throughout this letter we have assumed that the neutrinos move like photons in null geodesics, so the time delays do not depend on the energy of the neutrinos. The time lag between neutrinos with an energy $E_{\nu}$ and a small mass $m_{\nu}$ and photons travelling the same distance $d$ can be approximated by $\Delta t\approx (1/2)(d/c)(m_{\nu}c^2/E_{\nu})^2$, which using $E_{\nu} \gtrsim 1$ TeV (for neutrinos associated with GRBs), $m_{\nu}c^2 \lesssim 1$ eV, $d \sim 10^{28}$ cm,  gives $\Delta t < 10^{-6}$ s. This time lag corresponds to a total travelling time of about $3 \times 10^{17}$ s. Then, if neutrinos have mass, the time delays given in Tables \ref{table1} and \ref{table2}  should be modified in the same proportion, i.e. in less than one part in $10^{23}$. There is also observational evidence related to the supernova SN1987A which shows that the time delay due to the presence of the galaxy for photons and neutrinos with different energies is the same within a $0.5\%$ or better accuracy \cite{supernova}. All these justify our assumption, and the results obtained are excellent approximations if the neutrinos have sub-eV mass.\\

Concerning the relativistic images, if we choose $\beta /\theta_{\rm{E}}=0.1$ as it was done in Table \ref{table1}, we have from Eq. (\ref{pm9}) that the magnification of the brightest image is $\mu _{1}=1.1\times 10^{-17}$ for the Galactic black hole and $\mu _{1}=1.8\times 10^{-22}$ for NGC4486. Similar values are obtained for the other black holes considered in Tables \ref{table1} and \ref{table2}. The other image magnifications decrease exponentially with $n$. To obtain magnifications about one or larger, a closer alignment is necessary, i.e. $\beta \sim b_{\rm{ps}}/d_{\rm{ol}}$ instead of  $\beta \sim \theta _{\rm{E}}$. Then, while the primary and secondary images are amplified several times for $\beta /\theta_{\rm{E}}=0.1$, the strong deflection ones are highly demagnified. Using Eq. (\ref{pm7}), it can be seen that the angular separation between the strong deflection images is of the order of micro arc seconds or less. The time delays between the relativistic images, given by Eqs. (\ref{td3}) and (\ref{td4}), can be large, but they are too faint to be detected. So, in what follows, we restrict ourselves to the weak deflection images.\\

The probability of supermassive black holes located at the center of galaxies in the line of sight to GRBs is not negligible because of the high-redshift of most GRBs. Moreover, optical spectroscopic observations can detect absorbing lines of the interposed galaxy in the afterglow, hence allowing a direct determination of the different distances involved in the scenario. In the case of choked GRBs, where only neutrinos are produced, the time delays and the relative magnifications of the signals could used for the unequivocal identifications of {\sl dark} neutrino transients. A neutrino transient associated with a GRBs might have an fluence of several times $10^{-4}$ erg cm$^{-2}$ \cite{dermer}. With a mild amplification as obtained for the parameters adopted in Tables \ref{table1} and \ref{table2}, this might imply the detection of a few neutrinos by a km$^{3}$-detector. Even if the signal-to-noise ratio is not at a high confidence level, the repetition of the signal on a time scale from minutes to hours from the same location in the sky would render the identification of the neutrino transient source unequivocal. If the detection of the GRB afterglow allows a clear determination of the redshifts involved, then Eqs. (\ref{pm4}) and (\ref{td2}) can be used to obtain an independent estimated of the central black hole mass in the interposed galaxy. \\

The analysis of current databases indicates that the space-time clustering of GRBs is only marginal, at the level of $5\%$ or less \cite{romero}, but as we have mentioned in the Introduction, choked collapsars can result in transient neutrino sources without electromagnetic counterparts, so the total number of neutrino transients that is affected by lensing effects could be significantly larger from what is inferred from GRB population studies. The detection of a single event could be of paramount importance for our understanding of physical processes governing the GRBs.

\section{Final remarks}\label{conc} 

In this letter we have shown that the primary and secondary images of neutrino transient sources lensed by supermassive black holes cannot be angularly resolved but they could be temporally resolved by next generation instruments. The relativistic images, instead, are too faint and packed to be detected. Thus, we have found that neutrino transients produced by long GRBs can act as sources for gravitational lensing when supermassive black holes are present in foreground galaxies. This sources would have a unique signature, that will allow an easy detection above the background despite a possible low signal-to-noise ratio: repetition. The neutrino fluence can be magnified, but more importantly, the arriving signal will repeat, leading to an unequivocal identification. We conclude that neutrino gravitational lensing can help to establish GRBs as sources of relativistic protons and neutrinos, as proposed by several authors \cite{meszaros,dermer}.

\section*{Acknowledgments}

E.F.E. is supported by UBA and CONICET. G.E.R. is supported by grants PIP 5375 (CONICET) and PICT 03-13291 BID 1728/OC-AR (ANPCyT) and by the Ministerio de Educaci\'on y Ciencia (Spain) under grant AYA 2007-68034-C03-01, FEDER funds.


\begin{thebibliography}{99}

\bibitem{virbha1}  K.S. Virbhadra, and G.F.R. Ellis, Phys. Rev. D \textbf{62},
084003 (2000).

\bibitem{bozza1}  V. Bozza, S. Capozziello, G. Iovane, and G. Scarpetta, Gen. Relativ. Gravit. \textbf{33}, 1535 (2001).

\bibitem{schw} A.O. Petters, Mon. Not. R. Astron. Soc. \textbf{338}, 457 (2003); V. Bozza, and L. Mancini, Astrophys. J. \textbf{627}, 790 (2005); V. Bozza, and M. Sereno Phys. Rev. D \textbf{73}, 103004 (2006).

\bibitem {eiroto} E.F. Eiroa, G.E. Romero, and D.F. Torres, Phys. Rev. D
\textbf{66}, 024010 (2002).

\bibitem{bozza2} V. Bozza, Phys. Rev. D  \textbf{66}, 103001 (2002).

\bibitem{bozman1}  V. Bozza, and L. Mancini, Gen. Relativ. Gravit. \textbf{36}, 435 (2004).

\bibitem{bozza3}  V. Bozza, Phys. Rev. D  \textbf{67}, 103006 (2003); V. Bozza, F. De Luca, G. Scarpetta, and M. Sereno, Phys. Rev. D \textbf{72}, 083003 (2005); V. Bozza, F. De Luca, and G. Scarpetta, Phys. Rev. D \textbf{74}, 063001 (2006).

\bibitem{rotating} S.E. Vazquez, and E.P. Esteban, Nuovo Cimento B \textbf{119}, 489 (2004); W. Hasse, and V. Perlick, J. Math. Phys. (N.Y.) \textbf{47}, 024503 (2006).

\bibitem{alternative}  A. Bhadra, Phys. Rev. D  \textbf{67}, 103009 (2003); E.F. Eiroa, Phys. Rev. D \textbf{73}, 043002 (2006); K. Sarkar, and A. Bhadra, Class. Quantum Grav. \textbf{23}, 6101 (2006); N. Mukherjee, and A.S. Majumdar, Gen. Relativ. Gravit. \textbf{39}, 583 (2007); G. N. Gyulchev, and S. S. Yazadjiev, Phys. Rev. D \textbf{75}, 023006 (2007).

\bibitem{braneworld} V. Frolov, M. Snajdr, and D. Stojkovic,  Phys. Rev. D \textbf{68}, 044002 (2003); E.F. Eiroa, Phys. Rev. D \textbf{71}, 083010 (2005); R. Whisker, Phys. Rev. D \textbf{71}, 064004 (2005); A.S. Majumdar, and N. Mukherjee, Int. J. Mod. Phys. D \textbf{14}, 1095 (2005); E.F. Eiroa, Brazilian Journal of Physics \textbf{35}, 1113 (2005); C. R. Keeton, and A. O. Petters, Phys. Rev. D \textbf{73}, 104032 (2006).

\bibitem{nakedsing}  K.S. Virbhadra, D. Narasimha, and S.M. Chitre, Astron. Astrophys. \textbf{337}, 1 (1998); K.S. Virbhadra, and G.F.R. Ellis, Phys. Rev. D \textbf{65}, 103004 (2002);  K.S. Virbhadra, and C.R. Keeton, arXiv:0710.2333v1 [gr-qc] (2007).

\bibitem{darwin} C. Darwin, Proc. Roy. Soc London A \textbf{249}, 180 (1959).

\bibitem{old} J.-P. Luminet, Astron. Astrophys. \textbf{75}, 228 (1979); H.C. Ohanian, Am. J. Phys. \textbf{55}, 428 (1987); R.J. Nemiroff, Am. J. Phys. \textbf{61}, 619 (1993).

\bibitem{schneider} P. Schneider, J. Ehlers, and E.E. Falco, \textit{Gravitational Lenses} (Springer-Verlag, Berlin, 1992).

\bibitem{perturba} C.R. Keeton, and A.O. Petters, Phys. Rev. D  \textbf{72}, 104006 (2005); S.V. Iyer, and A.O. Petters, Gen. Relativ. Gravit. \textbf{39}, 1563 (2007).

\bibitem{amore} P. Amore, and S. Arceo Diaz, Phys. Rev. D  \textbf{73}, 083004 (2006); P. Amore, S. Arceo, and F.M. Fern\'{a}ndez, Phys. Rev. D  \textbf{74}, 083004 (2006); P. Amore, M. Cervantes, A. De Pace, and F.M. Fern\'{a}ndez, Phys. Rev. D  \textbf{75}, 083005 (2007).

\bibitem{retrolens}  D.E. Holtz, and J.A. Wheeler, Astrophys. J. \textbf{578}, 330
(2002); F. De Paolis, A. Geralico, G. Ingrosso, and A.A. Nucita, Astron. Astrophys. \textbf{409}, 809 (2003); E.F. Eiroa, and D.F. Torres, Phys. Rev. D \textbf{69}, 063004 (2004); F. De Paolis, A. Geralico, G. Ingrosso, A.A. Nucita, and A. Qadir, Astron. Astrophys. \textbf{415}, 1 (2004); V. Bozza, and L. Mancini, Astrophys. J. \textbf{611}, 1045 (2004).

\bibitem{bozza4} V. Bozza, and G. Scarpetta, Phys. Rev. D \textbf{76}, 083008 (2007).

\bibitem{escribano} R. Escribano, J.-M. Fr\`{e}re, D. Monderen, and V. Van Elewyck, Phys. Lett. B \textbf{512}, 8 (2001).

\bibitem{mena} O. Mena, I. Mocioiu, and C. Quigg, Astropart. Phys. \textbf{28}, 348 (2007).

\bibitem{neutrinos} M. Vietri, Astrophys. J. \textbf{507}, 40 (1998); J. Rachen, and P. M\'esz\'aros, Phys. Rev. D \textbf{58}, 123005 (1998); D. Guetta, M. Spada, and E. Waxman, Astrophys. J. \textbf{559}, 101 (2001).

\bibitem{meszaros} P. M\'esz\'aros, and E. Waxman, Phys. Rev. Lett. \textbf{87}, 171102  (2001).

\bibitem{dermer} C. D. Dermer, and A. Atoyan, Phys. Rev. Lett. \textbf{91}, 071102  (2003).

\bibitem{murase} K. Murase and S. Nagataki, Phys. Rev. D \textbf{73}, 063002 (2006); K. Murase and S. Nagataki, Phys. Rev. Lett. \textbf{97}, 051101 (2006); K. Murase, Phys. Rev. D \textbf{76}, 123001 (2007).

\bibitem{meszaros2} P. M\'esz\'aros, Rep. Prog. Phys. \textbf{69}, 2259 (2006).

\bibitem{woosley1} S.E. Woosley, Astrophys. J. \textbf{405}, 273 (1993).

\bibitem{woosley2} A.I. Macfayen and S.E. Woosley, Astrophys. J. \textbf{524}, 262 (1999).

\bibitem{atkinson} R. Atkinson, Astron J. \textbf{70}, 517 (1965).

\bibitem{claudel} C.M. Claudel, K.S. Virbhadra, and G.F.R. Ellis, J. Math. Phys. (N.Y.) \textbf{42}, 818 (2001).

\bibitem{lensequa} S. Frittelli and E.T. Newman, Phys. Rev. D \textbf{59}, 124001 (1999); S. Frittelli, T.P. Kling, and E.T. Newman, Phys. Rev. D \textbf{61}, 064021 (2000); V. Perlick, Phys. Rev. D \textbf{69}, 064017 (2004).

\bibitem{misner} C.W. Misner, K.S. Thorne, and J.A. Wheeler, \textit{Gravitation} (Freeman, New York, 1973).

\bibitem{richstone} D. Richstone \textit{et al.}, Nature \textbf{395}, A14 (1998).

\bibitem{romero} G.E. Romero, D.F. Torres, I. Andrichow, L.A. anchordoqui, and B. Link, Mon. Not. R. Astron. Soc. \textbf{308}, 799 (1999).

\bibitem{supernova} M. J. Longo, Phys. Rev. Lett. \textbf{60}, 173 (1988); L.M. Krauss and S. Tremaine, Phys. Rev. Lett. \textbf{60}, 176 (1988).

\end{thebibliography}
\end{document}